\documentclass[epj]{svjour}
\newcommand{\smartpap}{p\hskip-7pt\hbox{$^{^{(\!-\!)}}$}}
\usepackage{graphics}
\begin{document}
\title{Multi-photon corrections to $W$ boson mass determination 
at hadron colliders}
%\footnote{Presented with title ``Higher-order QED 
%corrections to $W$ boson mass determination 
%at hadron colliders''}
\author{C.M. Carloni Calame\inst{1} \and 
G. Montagna\inst{2} \thanks{\emph{Presented by G.~Montagna 
with title
``Higher-order QED corrections to $W$ mass determination
at hadron colliders''} }
\and 
O. Nicrosini\inst{1} \and M. Treccani\inst{3} % etc
}                     % Do not remove
%
%\offprints{G. Montagna, Dipartimento
%di Fisica Nucleare e Teorica, Universit\`a di Pavia,
%Via A. Bassi 6, I-27100 Pavia, Italy}          
% Insert a name or remove this line
%
\institute{Istituto Nazionale di Fisica Nucleare, Sezione di Pavia, 
%via A. Bassi 6, I-27100, Pavia, Italy 
and Dipartimento di Fisica Nucleare e Teorica, 
Universit\`a di Pavia, via A. Bassi 6, I-27100, Pavia, Italy
\and Dipartimento di Fisica Nucleare e Teorica, 
Universit\`a di Pavia, %via A. Bassi 6, I-27100, Pavia, Italy 
and Istituto Nazionale di Fisica Nucleare, Sezione di Pavia, 
via A. Bassi 6, I-27100, Pavia, Italy 
\and Dipartimento di Fisica Nucleare e Teorica, 
Universit\`a di Pavia, via A. Bassi 6, I-27100, Pavia, Italy}
\date{Received: date / Revised version: date}
\abstract{
The impact of higher-order final-state photonic corrections on the precise 
determination of the $W$-boson mass at the Tevatron and LHC
colliders is evaluated. The $W$-mass shift 
from a fit to the transverse mass distribution is found 
to be about 10~MeV in the $W \to \mu \nu$ channel and a few 
MeV in the $W \to e \nu$ channel. The calculation, which is implemented in 
the Monte Carlo event generator HORACE for data analysis, 
can contribute to reduce the uncertainty
associated to the $W$ mass measurement at present and future 
hadron collider experiments. 
\PACS{
      {12.15.Lk}{Electroweak radiative corrections}   \and
      {13.40.K}{Electromagnetic corrections to strong- and 
      weak-interaction processes}
     } % end of PACS codes
} %end of abstract
\maketitle
\section{Introduction}
\label{intro}
\begin{figure}
\resizebox{0.5\textwidth}{!}{%
  \includegraphics{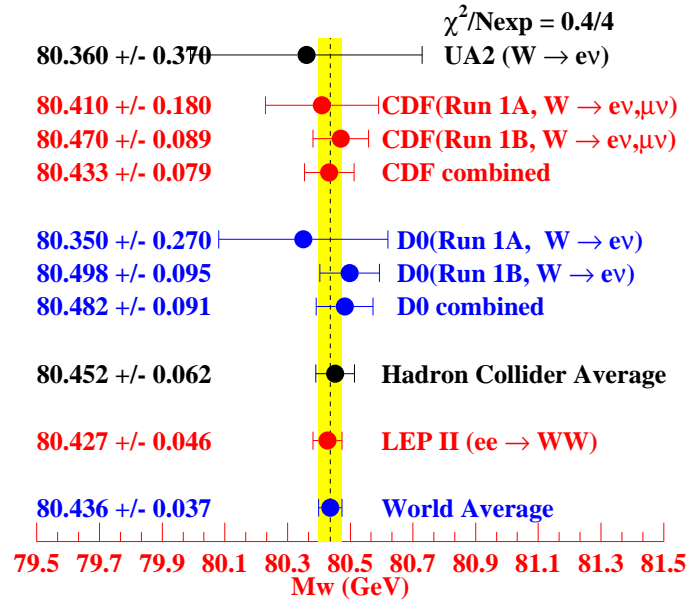}
}
%\vspace{5cm}       % Give the correct figure height in cm
\caption{Compilation of all the available direct 
measurements of the mass of the $W$ boson $M_W$.}
\label{fig:1}       % Give a unique label
\end{figure}

In addition to the program of discovery physics, experiments at the
high-energy hadron colliders Tevatron RunII and the LHC are expected to pursue
the program of precision physics successfully carried out during the last
decade at LEP, SLC and the Tevatron itself. In particular, for precision tests
of the Standard Model a very precise determination of the $W$ boson mass $M_W$
%at hadron colliders 
is of particular importance because this would give the possibility, 
when associated with an improved measurement of the top-quark mass, 
of putting more severe indirect bounds on the mass of 
the Higgs boson. The present experimental
situation for all the direct measurements of $M_W$ is shown in Fig.~\ref{fig:1}.
It is worth noticing that the 
hadron collider average for $M_W$ has an accuracy
comparable to the LEP measurement, but even more important is to take into account
that the precision expected at the Tevatron is about 30 MeV 
per experiment per channel at Run IIa and 16 MeV at Run IIb, 
the latter being the same precision aimed at at the LHC~\cite{exp,expb}. This 
competes with the precision of the $M_W$ measurement expected at a future 
$e^+ e^-$ collider and will correspond
to a knowledge of the $W$ mass with a relative precision of $2 \times 10^{-4}$, 
which obviously requires precise calculations and event generators for 
the Drell-Yan-like processes $p\smartpap \to W \to l \nu_l$
and $p\smartpap \to \gamma,Z \to l^+ l^-$, $l=e,\mu$.

%

%Your text comes here. Separate text sections with
\section{The $W$ mass at hadron colliders}
\label{sec:1}
At hadron colliders the $W$ mass is extracted from the two-body kinematics 
%features 
of the $W$ boson decay into a lepton and a neutrino $W \to l \nu_l$,
giving rise to a Jacobian peak in the distribution of the lepton transverse
momentum $p_T(l)$. However, the preferred quantity to determine $M_W$ 
is the transverse mass spectrum $M_T$, which is defined as
\begin{equation}
M_T = \sqrt{2 p_T(l) p_T(\nu) (1 - \cos\phi^{l\nu})}
\end{equation}
where $p_T(l)$ and $p_T(\nu)$ are the transverse momentum of the 
lepton and the neutrino, and $\phi^{l\nu}$ is the angle between the lepton 
and the neutrino in the transverse plane. Actually, the $M_T$ spectrum is less
sensitive than the $p_T(l)$ distribution to the transverse motion of the
$W$, which is difficult to model. Having in mind the precision anticipated at
the Tevatron and the LHC for $M_W$, accurate theoretical predictions including QCD and electroweak
radiative corrections are necessary to precisely extract $M_W$ from the data.
In the presently available calculations \cite{bkw,dk} $O(\alpha)$ electroweak corrections
are included exactly. From these calculations it comes out that electroweak corrections 
shift the $W$ mass by an amount of the order of $100$ MeV. In $W$ mass measurements at 
the Tevatron Run Ib the mass shifts due to radiative effects were estimated 
to be $-65 \pm 20$ MeV and $-168 \pm 10$ MeV for electron and muon channels, respectively
 \cite{exp,expb}. From previous studies, it is also known that
these shifts are mainly due to final-state photonic corrections because of the
presence of large collinear logarithms of the form 
$\alpha/\pi \log (\hat{s}/m_l^2)$,
where $\sqrt{\hat{s}}$ is the effective centre of mass (c.m.) energy 
and $m_l$ is the mass of the final-state lepton. Furthermore, the effect of
final-state QED corrections significantly depends on lepton identification requirements
and detector effects, which are different for electrons and muons. 
In the presence
of realistic selection criteria, 
the correction due to final-state photon radiation is of several per cent 
on the $M_T$ spectrum in the peak region $M_T \approx M_W$.
This poses the question of the impact of higher-order (i.e. beyond order $\alpha$)
QED corrections due to the multiple emission of (real and virtual) photons.
These higher-order contributions are not presently included in data analysis at 
the Tevatron but they are estimated to introduce a systematic uncertainty of
20 MeV in the $W \to e \nu_e$ decay channel and 10 MeV in the $W \to \mu \nu_\mu$ decay \cite{exp,expb}.
This source of systematic uncertainty is not negligible in view of the foreseen
experimental precisions and it can be reduced by means of improved theoretical calculations.
Recent work in this direction includes the calculation of the double bremsstrahlung
matrix elements $q \bar{q'} \to W \to l \nu \gamma \gamma$ 
and $q \bar{q} \to \gamma,Z \to l^+ l^- \gamma \gamma$ \cite{bs}, as well as the 
calculation of multi-photon corrections to leptonic $W$ decays in the
framework of Yennie-Frautschi-Suura exponentiation approach 
implemented in the Monte Carlo (MC) generator WINHAC \cite{winhac}.

\section{Theoretical approach and numerical results}
\label{sec:2}
In the approach here presented the real plus virtual corrections due to
multi-photon radiation are computed in the leading-log approximation using
the well-known and well-established QED structure-function approach. The
corrections are calculated by solving numerically the DGLAP equation for
the electron structure function by means of the QED Parton Shower algorithm
developed in Ref.~\cite{ps}. Only radiation from the final-state leptons 
is presently included in our approach (by simply attaching a single 
QED structure function to the final-state lepton), because it is known that quark-mass singularities,
originating from initial-state photon radiation, can be reabsorbed into a redefinition
of the Parton Distribution Functions (PDFs), in analogy to 
gluon emission in QCD \cite{sp}. After this mass-factorization procedure, initial-state-radiation
has only a small and uniform impact on the $M_T$ spectrum, while final-state radiation
significantly distorts the shape of the $M_T$ distribution, affecting
in turn the $M_W$ extraction. 
%Our procedure introduces a breaking of
%gauge invariance due to neglecting the radiation from the intermediate $W$
%boson, as well as the effect of the initial-final-state interference, but
%this breaking is at the level of non-logarithmic corrections, as can
%be checked by comparing with the exact results available in the literature.
 The formulation is implemented into a MC generator for data analysis,
HORACE (Higher Order RAdiative CorrEctions)~\cite{cmnt}. It is interfaced 
to PDFs and incorporates lepton identification criteria and detector
resolution effects, in order to perform simulations
for the hadronic process $p \smartpap \to W \to l \nu$, $l=e,\mu$ 
as realistic as possible. HORACE
can calculate photonic corrections to all orders and at order $\alpha$, 
to disentangle the effect of higher-order contributions and to compare with
the available $O(\alpha)$ programs. A first round of comparisons was performed
with WGRAD \cite{bkw} in Ref.~\cite{cmnt}, showing good agreement. 
A more detailed analysis, based on the comparison between HORACE and WINHAC, is in 
progress \cite{jpcmn}.

\begin{figure}
\resizebox{0.5\textwidth}{!}{%
  \includegraphics{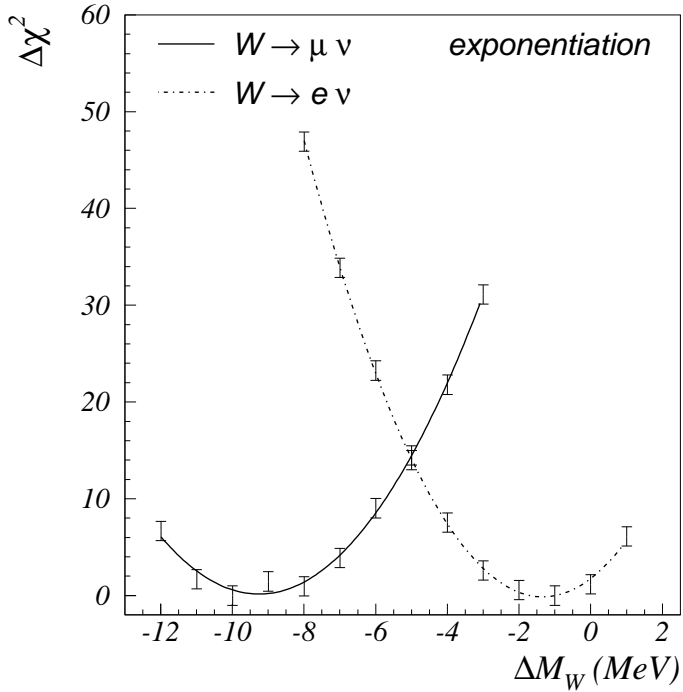}
}
%\vspace{5cm}       % Give the correct figure height in cm
\caption{The $\Delta\chi^2 = \chi^2 - \chi^2_{min}$ distributions 
from a fit to the $M_T$ spectrum as 
a function of the $W$ mass shift $\Delta M_W$ due to 
higher-order QED final-state corrections, at $\sqrt{s} = 2$~TeV.
 The results for the 
 $W \to e \nu$ and $W \to \mu \nu$ channels are shown.}
\label{fig:2}       % Give a unique label
\end{figure}

To evaluate the shift on the fitted $W$ mass induced by higher-order QED final-state
corrections we used HORACE and performed $\chi^2$ fits to MC pseudo-data 
for the $M_T$ spectrum, simulating acceptance cuts, lepton identification criteria and detector 
resolution effects as reported in Ref.~\cite{cmnt}. The c.m. energy considered
in our study is $\sqrt{s} = 2$ TeV, corresponding to the Tevatron Run II data taking, but we 
checked that the conclusions of our analysis do not change at the LHC energy
of $\sqrt{s} = 14$ TeV, when using the same cuts and detector specifications. 
The basic steps of the fitting procedure are the following: {\it i)} we generate a sample
of pseudo-data including $O(\alpha)$ corrections for a reference value of the
$W$ mass,  $M_W^{ref}$, and we simulate the $M_T$ spectrum within the fit
region $65~{\rm GeV} < M_T < 100$~GeV; {\it ii)} we consider $N$ different $W$-mass values around
$M_W^{ ref}$ and we generate $N$ $m_T$
spectra including higher-order corrections; {\it iii)} for each $M_W$ value, we calculate the
$\chi^2$ as
\begin{equation}
\chi^2 \, = \, \sum_i (\sigma_{i,exp} - \sigma_{i,\alpha})^2/
(\Delta\sigma_{i,exp}^2 + \Delta\sigma_{i,\alpha}^2)
\end{equation}
where $\sigma_{i,\alpha}$ and $\sigma_{i,exp}$ are the MC
predictions for the $i^{th}$ bin at the ${\cal O}(\alpha)$ and exponentiated level,
respectively, and
$\Delta\sigma_{i,\alpha}, \Delta\sigma_{i,exp}$ the corresponding
statistical errors due to numerical integration. We derive in conclusion the $M_W$ shift
due to higher-order effects looking at the minimum of the $\chi^2$ distribution. The 
results of our analysis are 
reported in Fig. \ref{fig:2}, showing the $\Delta\chi^2 = \chi^2 - \chi^2_{min}$ distributions 
as a function of $\Delta M_W \equiv M_W - M_W^{ref}$. It can be seen that the mass
shift due to higher-order effects is about $10$~MeV for the $W \to \mu \nu$ channel 
(solid line) and a few MeV (dashed line) for the $W \to e \nu$ channel, as a consequence of
the different identification requirements for electrons and muons. Therefore, in view of the
expected precision of 15-30 MeV for $M_W$ at the Tevatron Run II and at the LHC, it will
be important to take multi-photon effects into account when extracting $M_W$ from the data.

\section{Conclusions and outlook}
\label{sec:3}

In view of improved precision measurements of the $W$ mass at hadron colliders, 
we calculated higher-order QED final-state corrections to the process of single $W$ 
production in hadronic collisions. We found that the shift 
due to these corrections is about 10~MeV in the
$W \to \mu \nu$ channel and a few MeV in the $W \to e \nu$ channel.
The calculation, if included in future experimental analyses, 
would reduce the uncertainty in the precision measurement of the 
$W$ mass at hadron colliders. To this end, the Monte Carlo program HORACE is available 
for data analysis. We noticed that the shifts significantly depend on
particle identification requirements and detector effects. Therefore, a more precise 
assessment of our conclusions would require a full detector simulation, which needs for 
collaboration with experimental colleagues. Since $Z$ boson parameters, such as the 
$Z$ boson mass and width, are crucial
in determining the lepton energy and momentum scales, which enter and affect the
$M_W$ measurement, a calculation of higher-order corrections to the neutral-current process 
$p\smartpap \to \gamma,Z \to l^+ l^-$ is needed, 
being the $O(\alpha)$ electroweak corrections to the above process
already known in the literature~\cite{bz}. 
Work is in progress in this direction.

%

%
% BibTeX users please use
% \bibliographystyle{}
% \bibliography{}

\end{document}